\documentclass[pss,fleqn]{w-art}
\usepackage{graphicx}
\usepackage{amsmath}
\usepackage{epsfig}
\usepackage{w-thm}

\begin{document}

\setcounter{secnumdepth}{3}
\title{The geometry dependence of the conductance oscillations of monovalent
atomic chains}

\renewcommand{\copyrightyear}{2006}
\DOIsuffix{theDOIsuffix}
\Volume{XX} \Issue{} \Copyrightissue{} \Month{} \Year{}
\pagespan{1}{}
\Receiveddate{\sf } \Reviseddate{\sf } \Accepteddate{\sf } \Dateposted{\sf }
\subjclass[pacs]{73.23.-b, 73.23.Ad, 73.40.Cg, 73.63.-b, 73.63.Rt}

\author[P.~Major]{P{\'e}ter Major\footnote{Corresponding Author:
e-mail: {\sf majorp@complex.elte.hu}, Phone: +36\,20\,490\,2722,
Fax:+36\,1\,372\,28\,68}\inst{1,}\inst{2}}

\address[\inst{1}]{KFKI Research Institute of Particle and Nuclear Physics of  the
Hungarian Academy of Sciences, H-1121 Budapest, Konkoly-Thege u. 29-33,
Hungary}
\author[G.~Tichy]{G{\'e}za Tichy\inst{2}}
\address[\inst{2}]{Department of Solid State Physics, E{\"o}tv{\"o}s
University, H-1117 Budapest, P\'azm\'any P{\'e}ter s{\'e}t\'any
1/A, Hungary}

\author[J.~Cserti]{J{\'o}zsef Cserti\inst{3}}
\address[\inst{3}]{Department of Physics of Complex Systems, E{\"o}tv{\"o}s
University, H-1117 Budapest, P\'azm\'any P{\'e}ter s{\'e}t\'any
1/A, Hungary}

\author[V.~M.~Garc\'{\i}a-Su\'arez]{V\'{\i}ctor~Manuel Garc\'{\i}a-Su\'arez\inst{4}}

\address[\inst{4}]{Department of Physics, Lancaster University, Lancaster,
LA1 4YB, UK}

\author[S.~Sirichantaropass]{Skon~Sirichantaropass\inst{4}}

\begin{abstract}

Using a tight binding model we calculate the conductance of
monovalent atomic chains for different contact geometries. The
leads connected to the chains are modelled as semi-infinite fcc
lattices with different orientations and couplings. Our aim is
twofold: To check the validity of a three-parametric conductance
formula for differently oriented leads, and to investigate the
geometry dependence of the conductance oscillations. We show that
the character of these oscillations depends strongly on the
geometry of the chain-lead coupling.

\end{abstract}

\maketitle

\section{Introduction}  \label{intro:sec}

Since Ohnishi~\cite{Ohnishi:cikk} et al. and Yanson et al.~\cite{Yanson:cikk} demonstrated the existence of atomic
chains, there has been a rapidly growing interest in this field
[3--17]. The conductance of these chains was found to oscillate as
a function of length, as it had been predicted earlier in
Refs.~\cite{Pernas:cikk,Lang:cikk} and studied experimentally by e.g. Smit et al.~\cite{Smit:cikk}. This issue has
also been investigated theoretically for alkali metals, and it was
concluded that chains with odd numbers of atoms have a conductance
equal to $ 2e^2/h $, while the conductance is smaller than $ 2e^2/h $ for the chains with even numbers of atoms \cite{Sim:cikk, Gutierrez:cikk,
Tsukamoto_Hirose:cikk, Egami:cikk}, (hereafter we refer to this as
$ e<o $). The opposite of this, namely $ e>o $ was found in noble
metals~\cite{Lee:cikk} and also in some calculations of alkali
metals \cite{Lang:cikk}. The role of symmetries was also studied
in Refs.~\cite{Lee_Kim:cikk, Zeng:cikk} and the authors stated
that in chains with mirror symmetry and an odd numbers of atoms
the conductance is exactly $ 2e^2/h $. However, Havu et
al.~\cite{Havu:cikk} showed for sodium wires that the
oscillations may depend on the geometry. Hirose et
al.~\cite{Hirose_Kobayashi:cikk} also pointed out that the
oscillations are crucially influenced by the chain-lead coupling.

There exist different simple models that help to understand and
interpret the numerical and experimental results concerning the
conductance oscillations, such as in \cite{Smit:cikk, Sim:cikk,
Lee:cikk, Emberly:cikk, Lee_Sim:cikk}, but none of them is suitable
to explain the transformation between $ e<o $ and $ e>o $ due to
the geometry dependence.

\begin{vchfigure}[t]
\begin{center}
\includegraphics[scale=0.6]{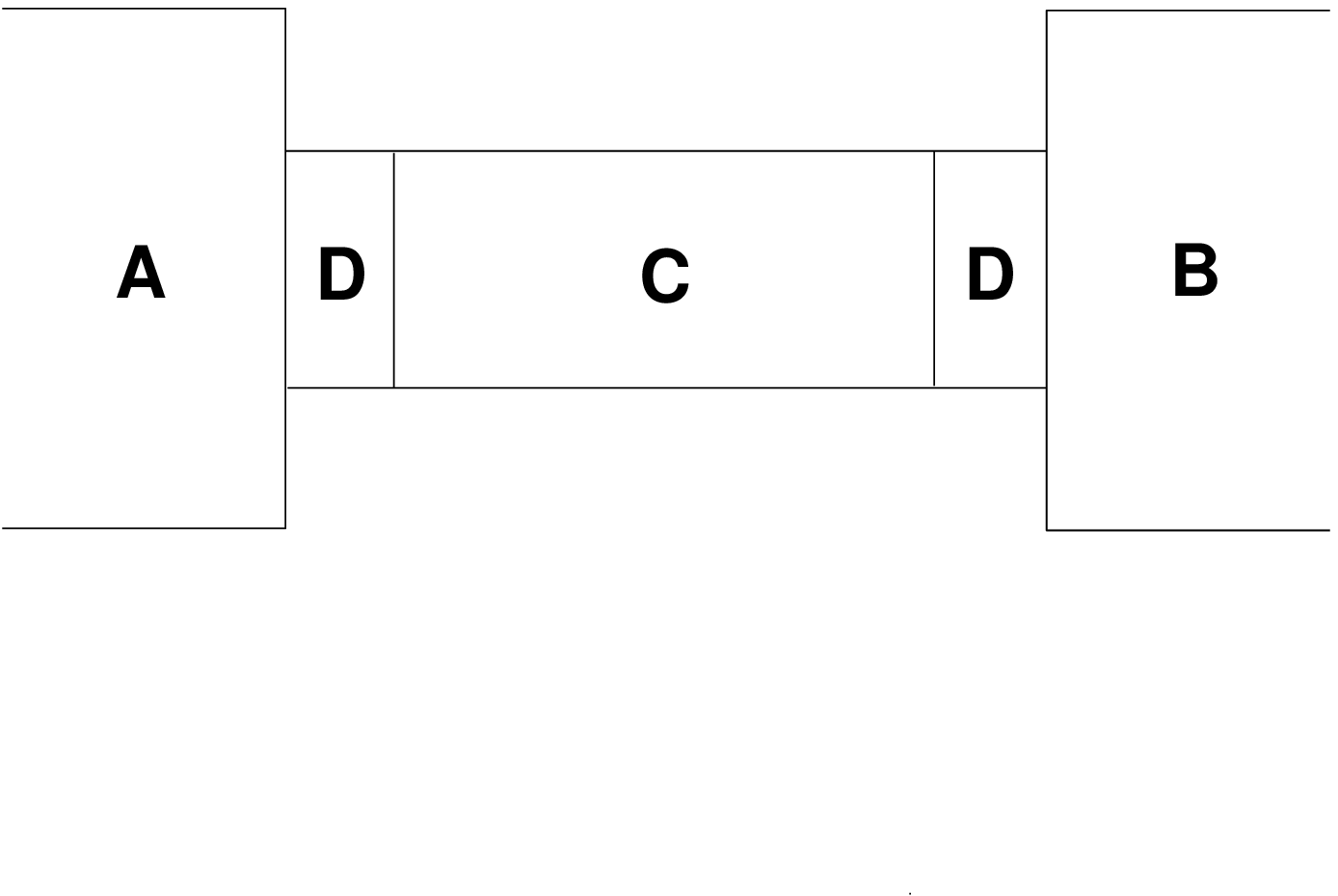}
\vchcaption{The lead(A)-chain(C)-lead(B) system. The coupling
atoms(D) between the chain and the leads are described by
scattering matrices with reflection coefficients $ r^{\prime}_{L}
$ and $ r_{R}$ those appear in Eq. (\ref{formula}).
\label{ballist}}
\end{center}
\end{vchfigure}

In our previous work~\cite{mi} we verified with the ab-initio code
SMEAGOL~\cite{Roc04,Natmat,Ptchains} that a three-parameter
formula can describe surprisingly well the conductance of
monovalent atomic chains, although the electron-electron
interaction could destroy its applicability. In units of $
2e^{2}/h $ the conductance as the function of length is given by

\begin{equation}
g =  \frac{\left(1-r^{\prime 2}_L \right)
\left(1-r^{2}_R \right)}
{1+ r_{L}^{\prime 2} r_{R}^2
- 2 r_{L}^{\prime} r_{R}
\cos \left(2kd + \Phi_{LR} \right)},
\label{formula}
\end{equation}
where $ d $ is the length of the translational invariant segment
of the chain, $ k $ is the wavenumber of the transmitting
electron, $ \Phi_{LR}=\Phi_{L}+\Phi_{R} $ and $ \Phi_{L}(\Phi_{R})
$ and $ r_{L}^{\prime} $ $ (r_{R}) $ are the phase shift and
magnitudes of reflection amplitudes at the left (right) side of
the chain. Eq.~(\ref{formula}) can be derived in the framework of
ballistic transport, using the Landauer-B\"uttiker
formula~\cite{Landauer-Buttiker}, considering two scattering
regions (namely the coupling atoms between the chain and the
leads), one conducting channel in the wire and a translational
invariant region within the atomic chain. The scheme of this
simple model can be seen in Fig.~(\ref{ballist}). In our previous
paper we investigated only symmetrical leads with one chosen
orientation on the two ends of the chain and we modified the
arrangement of the coupling atoms. However, in the experiments a
chosen lead orientation can not be guaranteed. Therefore it is
relevant to study whether the model can be applied to different
lead orientations and couplings. This is the aim of the present
paper.

In section \ref{model:sec} we describe the method we use to
calculate the conductance of atomic chains placed between fcc
leads. We explain how to obtain the Green's function of the leads,
the Green's function of the atomic chain and how to compute the
conductance of the total system. The numerical results are
presented in section \ref{num_res:sec}. We verify the validity of Eq.~(\ref{formula}) by fitting the results and obtaining the reflection coefficients and the phases for systems with various
leads orientations. If the ballistic model can be applied, then the fit parameters are not independent. For example by using two different symmetrical systems $\alpha\alpha$ and $\beta\beta$ (for the definitions see subsection \ref{final_S:sec}) it is possible to predict the conductance of an asymmetrical system $\alpha\beta$ with Eq.~(\ref{formula}). We compare this prediction with the result of the direct numerical calculation for the $\alpha \beta$ system. We finish with the conclusions.

\section{The model}
\label{model:sec}

To calculate the
differential conductance of an atomic chain placed between two
leads we use a tight-binding model developed by Todorov et.
al.~\cite{Todorov}. The Hamiltonian, which has identical on-site
energies and identical hopping energies reads

\begin{equation}
 \langle \mathbf{R}_{n}|\hat{H}|
\mathbf{R}_{m}\rangle = \left\{
\begin{array}{ll}
E_{0}, & \mbox{if $ \mathbf{R}_{n}=\mathbf{R}_{m} $ }, \\
\gamma ,  & \mbox{if $ \mathbf{R}_{n} $ and
$ \mathbf{R}_{m} $ are first
neighbours},  \\
0, & \mbox{otherwise} .
\end{array}
\right.
\label{eq:TB}
\end{equation}

where $ |\mathbf{R}_{n} \rangle $ is the atomic state of an electron located on the $n$th atom
with position vector $ \mathbf{R}_{n}$.
Hopping is allowed only between the first neighbors and the
topology of the atom arrangements determines the physical system.
In the calculations $E_{0}=0$ and $ \gamma=-1 $ is taken as the zero point 
and the unit of our energy scale. Three differently oriented blocks are investigated,
i.~e. the surface planes of the leads are (111), (100) and (110),
so we have six different lead pairs.

The two leads are formed from semi-infinite
stack of planes specified by their normal vector. 
The chain and the two leads are described by
the Hamiltonians $\hat{H}^s$ and $\hat{H}^{1,2}$, respectively,
and coupled by the potential $ \hat{V} $. The retarded Green's
function related to the total Hamiltonian $\hat{H}$ of the coupled
systems is given by $\hat{G}(E) = \lim_{\varepsilon \rightarrow
0^+}\, {\left(E- \hat{H} +i \varepsilon \right)}^{-1}$. To
calculate the conductance of the system, $ g $, we start from the
general expression derived by Todorov et
al.~\cite{Todorov}:

\begin{eqnarray}
g &=& \frac{4\pi e^{2}}{\hbar}\, {\rm Tr} \left[
\hat{\rho}_{1}(E_{F})\hat{t}^{\dag}(E_{F})
\hat{\rho}_{2}(E_{F})\hat{t}(E_{F})
\right],
\label{eq:conduct}
\end{eqnarray}
where $\hat{\rho}_{1,2} = -\frac{1}{\pi}\, {\rm Im }\,
\hat{G}^{1,2}$ are the density of states for lead $1$ and $2$
expressed via the corresponding Green's function $\hat{G}^{1,2}$
and the transfer operator $\hat{t}$ is defined by $\hat{t}(E) =
\hat{V}+\hat{V}\hat{G}(E)\hat{V}$. To calculate the conductance we
need then to determine the Green's functions $ \hat{G}^{1} $, $
\hat{G}^{2} $ and  $\hat{G}$ for the leads and for the coupled
system. The total Green's function $\hat{G}$ can be obtained from
the Dyson equation

\begin{equation}
\hat{G}=\hat{G_0} + \hat{G_0} \hat{V} \hat{G}, \label{Dyson:eq}
\end{equation}
where

\begin{equation}
\hat{G}_0=\hat{G}^1 +\hat{G}^2 + \hat{G}^s,
\end{equation}
and $\hat{G^s}$ is the Green's function of the isolated chain. In
the next two subsections we present the method we used to
calculate the Green's functions of the leads $\hat{G}^{1,2}$ and
the chain $\hat{G}^s$.

\subsection{The Green's function of the leads}
\label{Green_lead:sec}

An fcc crystal lead can be built up by stacking of its atomic
planes. The Bloch state of an electron in one of such planes is
given by

\begin{equation}
|l, \mathbf{q}\rangle = \frac{1}{\sqrt{N}}\sum_{n}
\exp(i\mathbf{q} \cdot \mathbf{R}_{n}^{l})|\mathbf{R}_{n}^{l} \rangle ,
\label{eq:basis}
\end{equation}
where  $ l $ denotes the $ l $th layer of the crystal, $\mathbf{q}
$ is the transverse wave vector on the $l $th layer and $
|\mathbf{R}^{l}_{n} \rangle $ is the atomic state of the electron located
on the $n$th atom of the $l$th layer (having position vector
$\mathbf{R}^{l}_{n} $). Here $1/\sqrt{N}$ is a normalization
factor.  The leads are supposed to be semi-infinite; $ l=0, 1, 2,
\dots \infty $. Both $ \lbrace | \mathbf{R}_{n}^{l} \rangle
\rbrace$ and $ \lbrace |l,\mathbf{q} \rangle \rbrace $ are
complete orthonormal bases. Every atom has $12$ nearest
neighbors. In the case of $(100)$ planes there are $4$
neighbors on the same plane while ($2 \times 4$) on the next two
adjacent planes. Similarly, for $(111)$ planes there are 6
neighbors located on the same plane and ($2 \times 3$) on the
adjacent planes. When the lead is built from $(110)$ planes there
are 2 neighbors on the same plane, and ($ 2 \times 4 $) on the
nearest and ($ 2 \times 1$) on the second nearest adjacent planes.
One can calculate the dispersion relation $ E(\mathbf{q}) $ of the
electrons on the layer as well as the coupling between layers,
i.e., the matrix elements $ \langle l, \mathbf{q}|\hat{H}|m,
\mathbf{q} \rangle $, where $ m=l, l \pm 1,l \pm 2 $ depending on
the orientation of the planes in the leads. To calculate the
density of states that appears in Eq.~(\ref{eq:conduct}) we need to
construct the Green's functions $\hat{G}^{1,2}$ of the leads from
the solution of the Dyson equation for the semi-infinite stack of the
coupled planes. In the case of the orientation $(111)$, the
Green's function has already been derived by Todorov et
al.~\cite{Todorov}. In a similar way one can derive an explicit
expression for the Green's function in the case of stacking of
$(100)$ planes. However, for stacking of $(110)$ planes the method
used in Ref.~\cite{Todorov} cannot be applied because every atom
has neighbors not only on the first but also on the second
adjacent layers. In this case we used Haydock recursion~\cite{hay}
to calculate the Green's function numerically. The smoothing of
the density of states is achieved by the commonly used termination
of the recursion. The chain is connected to the leads in such a
way, that the position of the first atom of the chain is equal to
the atomic position in the next imaginary atomic plain of the
lead. (Index $l$ of this fictional plain would be equal to -1.) In the (111), (100) and (110) orientations there are 3, 4
and 5 connecting atoms, respectively.

\subsection{The Green's function of the atomic chain}
\label{Green_chain:sec}

We now consider a one-dimensional atomic chain with $n$ atoms
labelled by $s_1, s_2, \cdots, s_n$. In the tight-binding
approximation the Hamiltonian of the chain can be written as
\begin{equation}
\hat{H}^s = E_0 \sum_{i=1}^n \, |s_i\rangle \langle s_i|
+ \gamma \sum_{i=1}^{n-1} \, \left( |s_i\rangle \langle s_{i+1}|
+ |s_{i+1}\rangle \langle s_i|  \right),
\end{equation}
where $\{|s_i\rangle \}$ are the orthonormal atomic basis in the
chain. Due to the special form of $\hat{H}^s$, the Green's
function $\hat{G}^s (E) = \lim_{\varepsilon \rightarrow
0^+}\,{\left(E-\hat{H}^s +i \varepsilon \right)}^{-1}$ can easily
be calculated using  recursively the following identity:
\begin{equation}
{\Bigl(\, A+
| x \rangle
\langle y | \,
\Bigr)}^{-1} = A^{-1} -
\frac{A^{-1} | x \rangle \langle y | A^{-1}}{1+\langle y |
A^{-1} | x \rangle }.
\label{azonossag}
\end{equation}
This identity is valid for arbitrary vectors $| x \rangle $ and $|
y \rangle $. This method is more versatile than the one used in
Ref.~\cite{Todorov}. It can be used, e.~g. for atomic chains with
dangling segments or loops of atoms.

\subsection{The conductance}
\label{final_S:sec}

The conductance given by Eq.~(\ref{eq:conduct}) can be obtained by
performing the trace in the atomic basis $\lbrace |\mathbf{R}_{n}
\rangle \rbrace$. After straightforward manipulation we find
\begin{equation}
g_{\alpha,\beta} = \frac{8  e^{2}}{h} \,
 {\rm Im}\, (f_{\alpha}) \, {\rm Im}\, (f_{\beta}) \,
|\langle s_{1}|\hat{G}(E_{F})|s_{n} \rangle |^{2}\, |\gamma|^{4},
\label{conduct_final:eq}
\end{equation}
where $\alpha$ and $\beta$ denote the orientation of the leads $1$
and $2$, i.~e. $(100)$, $(110)$ or $(111)$ planes. Factors
$f_\alpha$ and $f_\beta$ in Eq.~(\ref{conduct_final:eq}) are
related to the non-zero matrix elements of the density of states
of the leads.  For the three different orientations we find the
following explicit forms:
\begin{subequations}
\begin{eqnarray}
f_{111} &=&\frac{1}{N^{2}} \sum_{i,j=1}^{N}\,
G_{00}^{(111)}\, \Biggl(
3+6 \cos \frac{2\pi i}{N} \Biggr),
\label{eq:111fact}  \\
f_{100} &=& \frac{4}{N^{2}}\sum_{i,j=1}^{N} \,
 G_{00}^{(100)} \, \Biggl(
1+ \cos\frac{2\pi i}{N} + \cos\frac{2\pi j}{N} + \cos \frac{2\pi i}{N} \cos \frac{2\pi j}{N} \Biggr),  \\
\label{eq:100fact}
f_{110} &=& \frac{4}{N^{2}} \sum_{i,j=1}^{N} \left[
G_{00}^{(110)} \Biggl( 1+ \cos \frac{2\pi i}{N} +\cos \frac{2\pi  j}{N}
+\cos \frac{2\pi i}{N} \cos \frac{2\pi j}{N} \Biggr)
\right. \nonumber \\
&& \left. +\frac{G_{11}^{(110)}}{4} +\bigl( G_{01}^{(110)} + G_{10}^{(110)} \bigr)\cos \frac{\pi i}{N} \cos \frac{\pi j}{N} \right] ,
\label{eq:110fact}
\end{eqnarray}
\label{f_alpha:def}
\end{subequations}
where $ G_{l,m}^{\alpha} $ are the matrix elements of the Green's
function between layers $l$ and $m$ of the $\alpha$ oriented lead. Finally, in
Eq.~(\ref{conduct_final:eq}) $\langle
s_{1}|\hat{G}(E_{F})|s_{n}\rangle$ are the matrix elements of the
total Green's function of the system between the first and the
last atoms of the chain. Using the Dyson equation (\ref{Dyson:eq}) we
obtain
\begin{equation}
\langle s_{1}|\hat{G}(E_{F})|s_{n}\rangle = \frac{G^{s}_{1n}}{1-(G^{s}_{11}f_{\alpha}-G^{s}_{nn}
f_{\beta})|\gamma|^{2}+D(G^{s})f_{\alpha}f_{\beta}|\gamma|^{4} },
\end{equation}
where $ D(G^{s})=\big(G^{s}_{11}G^{s}_{nn}-G^{s}_{1n}G^{s}_{1n}
\big) $ with matrix elements $G^{s}_{ij}=\langle
s_{i}|\hat{G}^{s}| s_{j} \rangle $ of the Green's function for the
atomic chain. We now have all the necessary formulas to calculate
numerically the conductance.

\section{Numerical results}
\label{num_res:sec}

The dispersion relation of the chain in tight-binding approximation reads
\begin{equation}
E(k) = E_{0} + 2  \gamma
\cos(k  a),
\label{disp}
\end{equation}
where $a$ is the lattice constant. Thus, the chain has states only
in the $E_{0}-2\gamma \dots E_{0}+2\gamma$ interval therefore this is the energy
range where the system can work as a conductor. We have taken 9
different energy points in this interval and calculated the
conductance with 6 different lead pairs. As a first step we checked the validity of the ballistic transport model by doing
some tests. We fitted Eq.~(\ref{formula}) on the numerical results to
check whether the required relations between the parameters are
fulfilled or not. For the phase $ \Phi_{LR} $ in
Eq.~(\ref{formula}) there holds an addition rule. The phase of the
conductance oscillations of a chain between $\alpha$ and $\beta$
leads is related to the phases of the oscillations between the
symmetrical ($\alpha \alpha$ and $\beta \beta$) leads as

\begin{equation}
\label{phase_add}
\Phi _{\alpha\beta} =
\frac{\Phi_{\alpha\alpha}+\Phi _{\beta\beta}}{2}.
\end{equation}
This relation is fulfilled by our numerical results with an
accuracy better than 1\%. The values of the phases from the fit of
Eq.~(\ref{formula}) are summarized in Table~\ref{tab}. The good
agreement is clearly visible at every energy level. The energy
dependence of the phases can be seen in Fig.~\ref{fazisok}. For
symmetrical leads $ r '_{L} = r_{R} $, so these parameters
can be determined also from the fit as it can be seen in
Table~\ref{tab} and in Fig.~\ref{reflex}.

\begin{vchtable}[b]
\renewcommand{\arraystretch}{1.5}
\begin{center}
\vchcaption{\label{tab}The phase $ \Phi $ and reflection amplitude
$ r $ of the oscillations at different energies, resulted from the
fit of Eq.~(\ref{formula}).}
\begin{tabular}{c|cccccc|ccc}
$ E [\gamma] $ & $ \Phi_{\mathrm{111,111}} $ & $ \Phi_{\mathrm{100,100}} $ &  $ \Phi_{\mathrm{110,110}} $ & $ \Phi_{\mathrm{111,100}} $ & $ \Phi_{\mathrm{111,110}} $ & $ \Phi_{\mathrm{100,110}} $ & $ r_{\mathrm{111,111}} $ & $ r_{\mathrm{100,100}} $ & $ r_{\mathrm{110,110}} $ \\
\hline
-1.96 & -2.976 & -2.741 & -2.591 & -2.861 & -2.781 & -2.661 & 0.651 & 0.706 & 0.712 \\
-1.5 & -2.386 & -1.551 & -1.047 & -1.968 & -1.716 & -1.299 & 0.199 & 0.314 & 0.368 \\
-1.02 & -1.300 & -0.626 & -0.151 & -0.972 & -0.745 & -0.391 & 0.085 & 0.233 & 0.311 \\
-0.92 & -0.955 & -0.476 & 0.024 & -0.696 & -0.446 & -0.226 & 0.071 & 0.224 & 0.309 \\
-0.5 & 1.101 & 0.227 & 0.547 & 0.660 & 0.810 & 0.395 & 0.067 & 0.206 & 0.307  \\
0.02 & 2.336 & 1.004 & 1.093 & 1.667 & 1.718 & 1.042 & 0.114 & 0.213 & 0.323 \\
0.5 & 2.787 & 1.605 & 1.515 & 2.200 & 2.140 & 1.560 & 0.177 & 0.244 & 0.353 \\
1.02 & 3.033 & 2.090 & 1.985 & 2.615 & 2.440 & 2.070 & 0.272 & 0.315 & 0.415  \\1.6 & 3.124 & 2.629 & 2.434 & 2.869 & 2.779 & 2.544 & 0.456 & 0.479 & 0.559
\end{tabular}
\end{center}
\end{vchtable}

\begin{vchfigure}
\begin{center}
\includegraphics[scale=0.55]{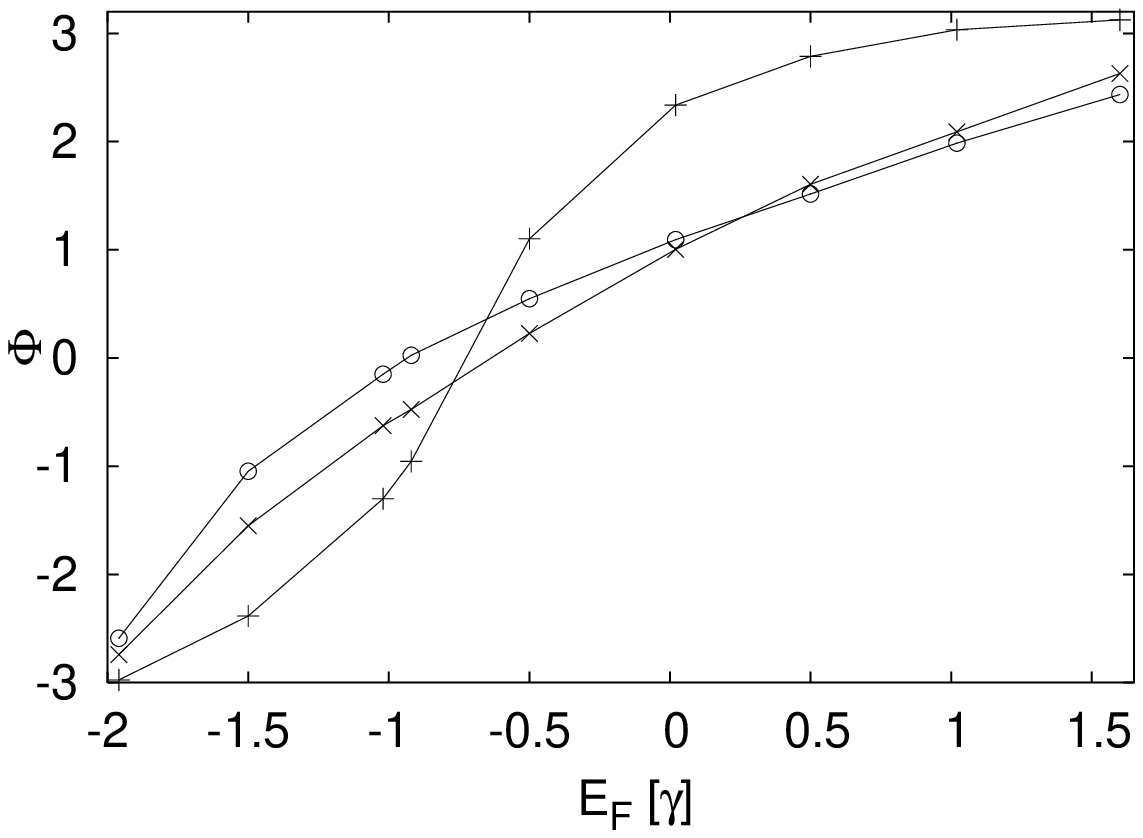}
\vchcaption{The energy dependence of the phases $\Phi_{LR}$ for symmetrical leads. Signs $ + $, $ \times $ and $ \circ $ denote the orientations (111), (100) and (110), respectively.
\label{fazisok} }
\end{center}
\end{vchfigure}

\begin{vchfigure}
\begin{center}
\includegraphics[scale=0.55]{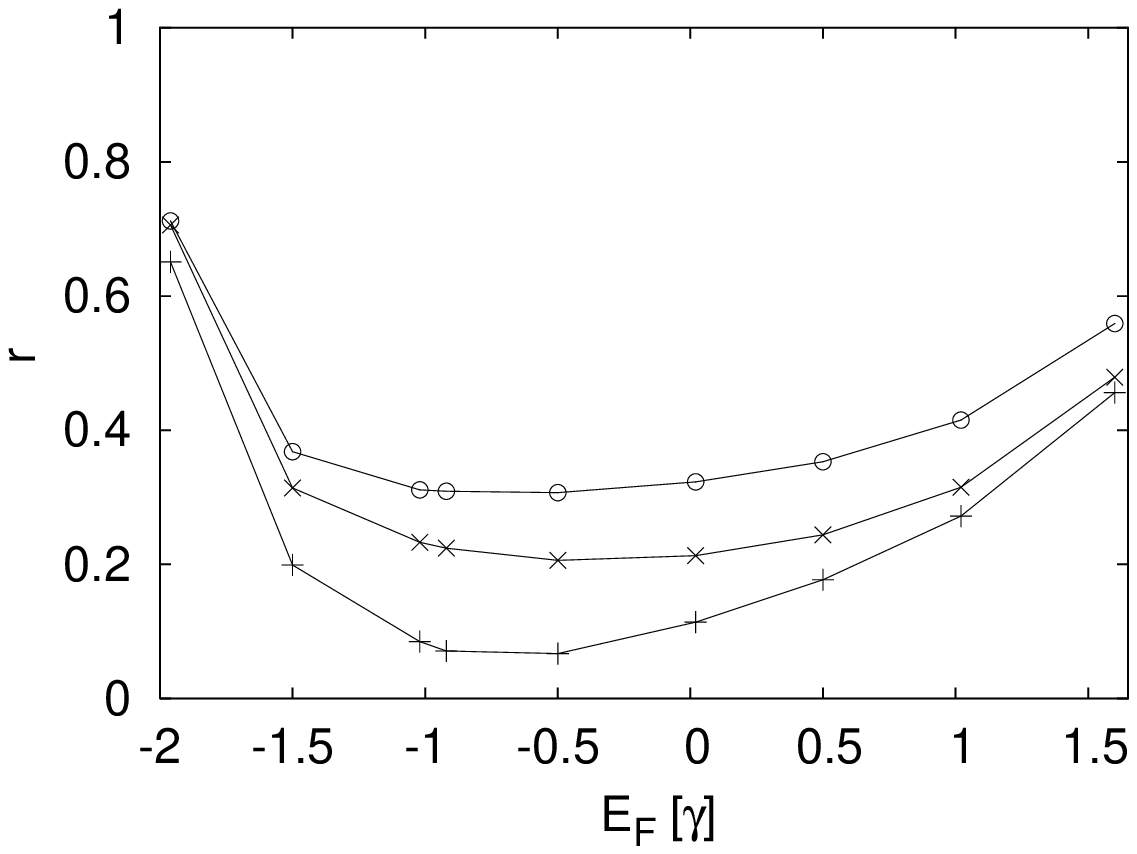}
\caption{The energy dependence of the reflection coefficients $r$ for symmetrical leads. Signs $ + $, $ \times $ and $ \circ $ denote the orientations (111), (100) and (110), respectively.
\label{reflex} }
\end{center}
\end{vchfigure}


Another proof of the validity of Eq.~(\ref{formula}) can be
obtained as follows: By calculating the conductance of two
different but symmetrical systems, for example between
$\alpha\alpha$ and $\beta\beta $ leads, with the fitted
parameters, $ \Phi_{\alpha \alpha}, \Phi_{\beta \beta},
r_{\alpha}, r_{\beta}$, the conductance of the asymmetric
$\alpha\beta$ system can be predicted. To do this we need to
substitute $ \Phi_{\alpha \beta} $ from Eq.~(\ref{phase_add}), $
r_{L}'=r_{\alpha} $ and $ r_{R}=r_{\beta} $ in Eq.
(\ref{formula}). The comparison of the predicted and the directly calculated results is plotted in Fig.~\ref{check2} and
implies that differently oriented leads do not modify the
availability of Eq.~(\ref{formula}). The orientation of the leads
influences the conductance essentially due to the different
chain-lead coupling and different density of the atoms on the
surface. The above results suggest that all of these properties
can be described by the parameters $r'_{L}$, $r_{R}$ and
$\Phi_{LR} $, and we do not need to modify the simple ballistic
transport model for different leads.

\begin{vchfigure}[t]
\begin{center}
\includegraphics[scale=0.25]{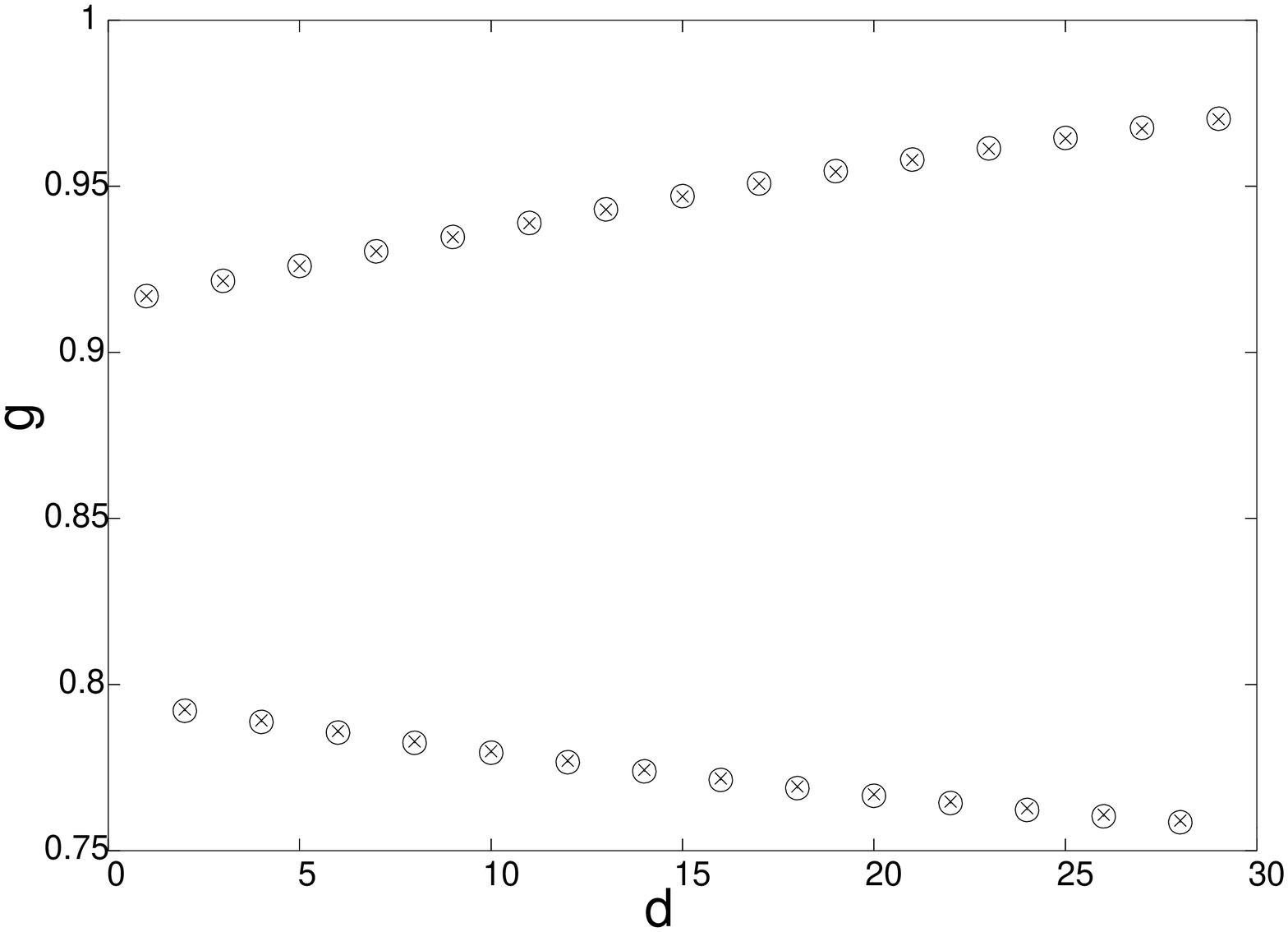}
\vchcaption{The conductance (in units of $2e^2/h$) of a chain
between asymmetric ($\alpha \beta $) leads, at energy
$E_{F}=0.02\gamma $ determined by Eq.~(\ref{formula}),
substituting the parameters of the symmetrical $\alpha \alpha$ and
$\beta \beta$ systems (crosses), and that from direct calculation
(open circles) as a function of the length $d$ (in units of the
lattice constant $a$) for lead orientations $\alpha = (110)$ and
$\beta = (100)$. \label{check2}}
\end{center}
\end{vchfigure}

\begin{figure*}
\begin{tabular}{ccc}
\includegraphics[scale=0.37]{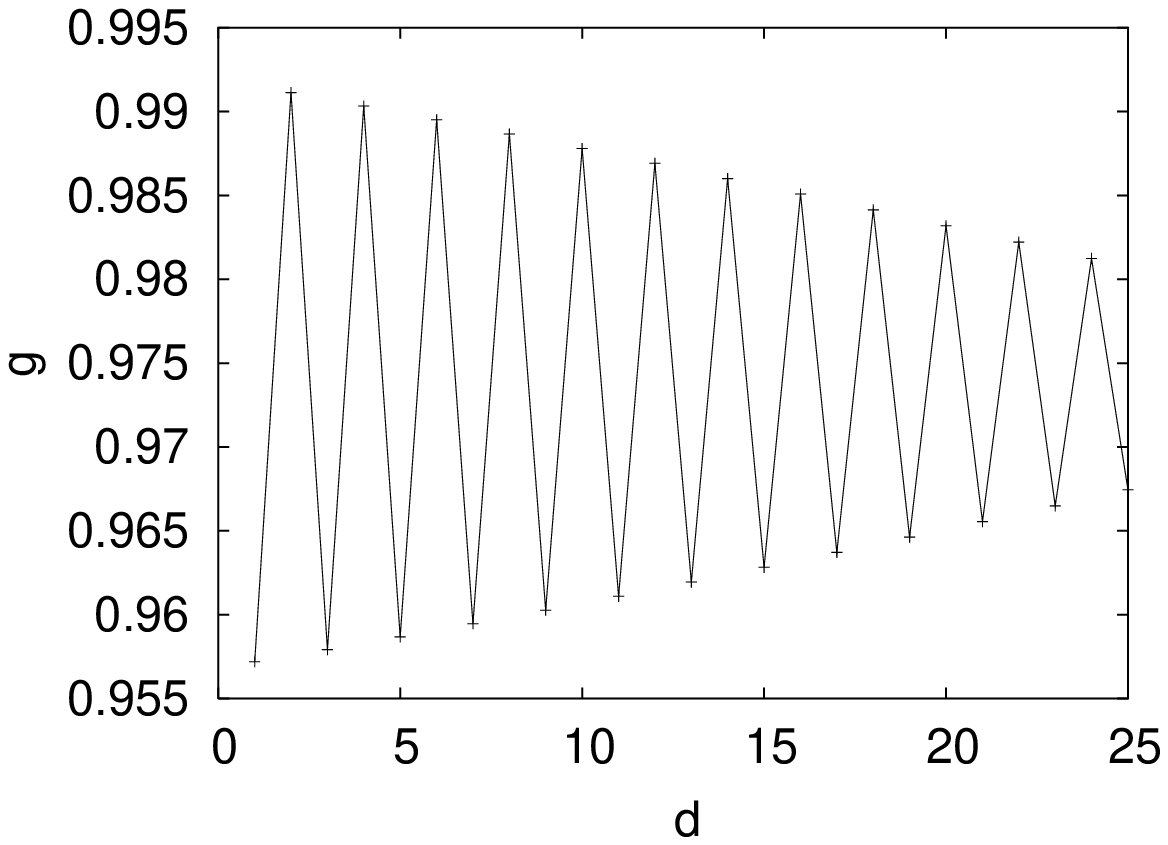}
&
\includegraphics[scale=0.37]{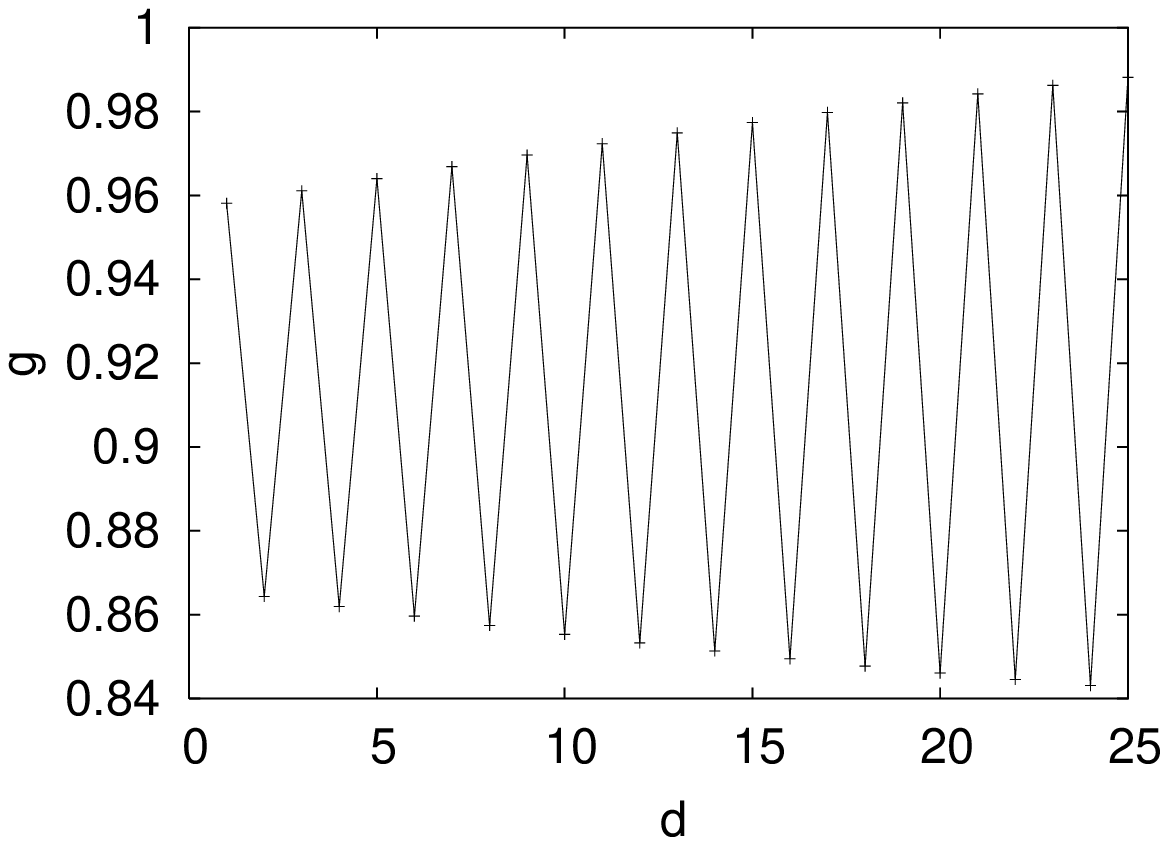}
&
\includegraphics[scale=0.37]{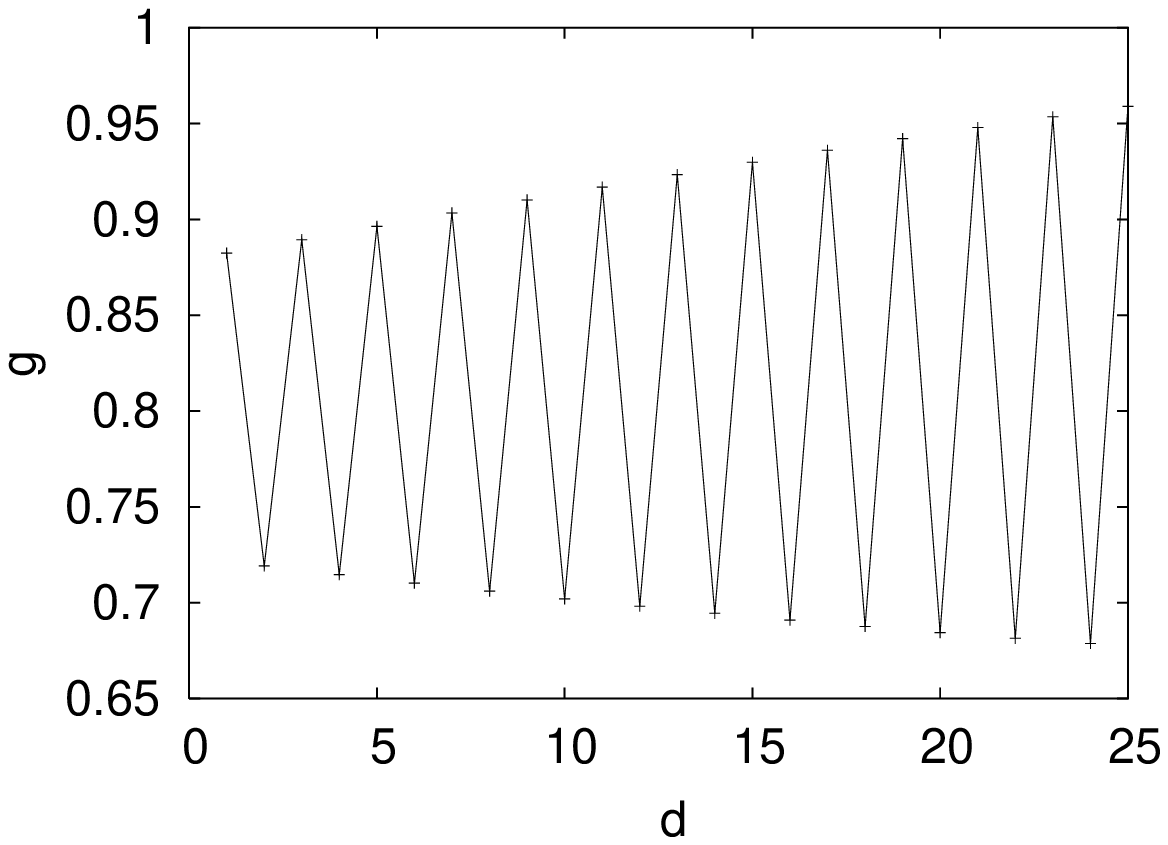} \\

a & b & c
\end{tabular}
\caption{\label{harom} The conductance (in units of $2e^2/h$) as a
function of the length $d$ (in units of the lattice constant $a$)
of the chain between symmetrical leads at $E_F = 0.02 \gamma$.
Figs. a, b and c show the results for the lead orientations
111-111, 100-100 and 110-110, respectively.}
\end{figure*}

We found that the wavelength of the oscillations does not depend on
the length and the character of the leads. It is essentially
determined by the dispersion relation of the chain, similarly to
our earlier results~\cite{mi}. At half filled band the wavenumber
is $ \pi/(2a) $, which gives two-atom-periodicity, but at other Fermi
levels it becomes different. For example, for wavenumbers $
k=\pi/(4a) $ or $ k=(3\pi)/(4a) $ it yields four-atom-periodic
oscillations, in agreement with the results of Thygesen and
Jacobsen~\cite{Thygesen:cikk}, who found that the three
conductance channels of aluminium have wavenumbers $\pi/(4a) $, $
\pi/(2a) $, and $ (3\pi)/(4a)$, and the conductance oscillations
have four atoms periodicity. The periodicity of the oscillations
given by Eq.~(\ref{formula}) is also in agreement with the earlier
prediction of de la Vega and co-workers~\cite{de_la_Vega}, who
predicted that every conduction channel oscillates as a function
of length with a wavenumber $ 2k_{F} $.

For monovalent atomic chains the charge neutrality condition
requires wavenumber $ k_{F} =\pi/(2a)$. Then, it follows
from Eq.~(\ref{formula}) that the conductance is an alternating
function of the number of atoms. At $ k_{F}=\pi/(2a)$ the phase
can not be determined easily from the fit on numerical data and it
seems to be an unnecessary parameter. Maybe this fact can explain
why the phase was missed in earlier interpretation
models~\cite{Smit:cikk, Sim:cikk, Lee:cikk, Emberly:cikk,
Lee_Sim:cikk}. However, in spite of these difficulties the phase is
very useful to interpret the geometry dependence of the
oscillations. To avoid numerical problems, we could take the
limits $ \lim_{E \rightarrow E_{F}}\,\Phi(E) $ and $ \lim_{E
\rightarrow E_{F}}\,r(E) $, but as can be seen in
Figs.~\ref{fazisok} and~\ref{reflex} these parameters are
smooth functions of the energy and therefore we choose $ E_F=0.02
\gamma $ in the calculations. For symmetrical leads some numerical
results are plotted in Fig.~\ref{harom}. Because of the
different groups of atoms connecting the chain to the lead, the
oscillations have different phases. Between leads with planes
(100) and (110) one can see that the conductance for chains with
an even atoms is smaller than for chains with an odd number of
atoms, i.~e. $ e<o $. However, in the case of (111) orientation, the so called
even-odd effect changes its sign and $ e>o $. This demonstrates
the sensitive geometry dependence of the conductance. The analysis
of Eq.~(\ref{formula}) implies that the maximum value of the conductance can be $
(2e^{2}/h) $ only if the cosine term is equal to 1 and therefore
the phase is 0. It is also clear from the figure that not only the
phases, but also the amplitudes depend on the coupling, since
according to Eq.~(\ref{formula}) larger reflection coefficient $r$
yields larger amplitude of the oscillation. Indeed we have found
the relation $ r_{111} < r_{100} < r_{110}$, (0.114, 0.213 and 0.323, respectively at $E_{F}=0202\gamma$) which follows the same
pattern as the amplitudes of the oscillations. In the case of
orientations (111), (100), (110) there are 3, 4 and 5 coupling
atoms between the lead and the chain, respectively, suggesting
that more atoms may increase the reflection amplitudes. However,
it is also necessary to stress that the different chain-lead couplings
result in different phases, and these can modify the amplitudes of
the oscillations also.

\section{Conclusions}
\label{summary:sec}

We have shown that the differently oriented leads do not affect
the applicability of Eq.~(\ref{formula}). However, the geometrical
arrangement of the atoms connecting the chain to the lead appreciably
influences the conductance, since the phases and the amplitudes of
the oscillations change significantly and the so-called even-odd effect can change its sign. 
This implies that numerical results of different
calculation methods are comparable only in the case of equal
geometrical circumstances. The validity of Eq.~(\ref{formula}) can
advance further investigations of the geometry dependence of the
oscillations, since it can be used to predict the transport
characteristics of asymmetrical leads, so that it would only be
necessary to perform simulations of symmetrical systems.

Since the parameters of the oscillations depend so sensitively on
the details of the coupling between the chain and the leads, it is
not clear  at first sight why the even-odd effect can be detected
experimentally. A possible explanation is the following. In the
experiments the average of conductances of several
different contacts is measured. The different geometries and consequently
oscillations with different phases could suppress the even-odd
effect. However, there are energetically preferred
surface orientations for a chosen material, and these facets
assist the formation of preferred chain-lead couplings, resulting
favored certain $r$ and $\Phi$ parameter values. Rodrigues et al.~\cite{Rodrigues1, Rodrigues2} investigated the effect of
these preferred surfaces on the conductance of point contacts
experimentally, and they found that in the case of gold the (111)
facets are more favorable. An other argument can be the following.
In the experiments after one breaking of the contact the tips are
brought together again to form a new contact. The displacement and
variation of both parts are usually small. Therefore the recreated
contact has similar geometry and presumably similar oscillation
parameters.

In summary we showed that Eq.~(\ref{formula}) and specially its
phase plays an important role to interpret the conductance
oscillations of atomic wires of monovalent elements.

\begin{acknowledgement}

The authors would like to thank the helpful discussions with
Gergely Szirmai and Tam\'as Herpay. VMGS thanks the European Union
for a Marie Curie grant.
\end{acknowledgement}


\end{document}